# Thermal properties of SnSe nanoflakes by AFM-based Scanning Thermal Microscopy measurements

M. Ozdogan[1,‡], T. Iken[1,‡], D. Cakir[1], N. Oncel[1,*]

*Department of Physics and Astrophysics, University of North Dakota, Grand Forks, ND 58202, United States of America*

(*corresponding author: nuri.oncel@und.edu);

[‡]Equal contribution

**Abstract:**

We investigated the thermal resistance of SnSe nanoflakes using scanning thermal microscopy, an extension of atomic force microscopy that allows for nanoscale thermal mapping. The nanoflakes studied ranged in thickness from 5 nm to 40 nm, enabling us to observe the thermal properties across a wide range of dimensions. Our results show a clear trend: as the thickness of the nanoflakes increases, the thermal resistance also increases. This suggests that diffusive mechanisms govern heat transport in SnSe nanoflakes and follow Fourier's law of heat conduction. The observed behavior can be attributed to the inherently short phonon mean free path in SnSe, which is on the order of 5 nm at room temperature.

## Introduction:

Over 50% of the energy produced globally is dissipated as waste heat,[1] representing a substantial loss of potential resources. This significant amount of energy can be recovered through thermoelectric materials, which can convert waste heat into electricity by utilizing the Seebeck effect. The efficiency of a thermoelectric material in converting heat to electricity is quantified by a temperature-dependent, dimensionless figure of merit, $zT$, defined as $zT = \sigma S^2 T / \kappa$.[2] In this equation, $\sigma$ represents the electrical conductivity, S is the Seebeck coefficient, and $\kappa$ denotes the total thermal conductivity, including contributions from both the movement of free electrons ($\kappa_{el.}$) and lattice vibrations (phonons, $\kappa_{lat.}$). Consequently, optimal thermoelectric materials require high electrical conductivity, a large Seebeck coefficient, and low thermal conductivity.

While 2D materials such as graphene, [3–5] black phosphorus, [6] and transition metal dichalcogenides (TMDCs) [7,8] have received more attention than SnSe, the remarkable thermoelectric performance of SnSe warrants deeper exploration. Its outstanding figure of merit underscores its potential for significant advancements in thermoelectric applications. SnSe features a layered orthorhombic crystal structure, [9] having approximately 0.6 nm thick layers,[10] strong in-plane bonds along the b–c axes while weaker Sn–Se bonds form along the a-axis,[11] facilitating easy cleavage in this direction (Figure 1a). It is worth noting that the bonding between the layers is somewhat unique.[9] While the bonding between the layers of SnSe is weak and van der Waals-like, it is not purely van der Waals. Small contributions from residual covalent interactions make the interlayer bonding slightly stronger than in purely van der Waals materials. The anisotropic structure and unique bonding nature of SnSe lead to directional variations in electrical and thermal conductivities.[11] By taking advantage of this, Zhao et al. reported a record $zT$ value of 2.6 along the b-axis in single crystal bulk SnSe.[12] Additionally, Zhou et al. enhanced this value to 3.1 in hole-doped polycrystalline SnSe samples by reducing the thermal conductivity by an order of magnitude.[13] SnSe has been extensively studied in its bulk form;[14,15] however, exploring its thermoelectric properties at the few-layer (or 2D) level can offer new insights and help designing more efficient thermoelectric devices. It is well-accepted that nanostructuring enhances the efficiency of thermoelectric materials. Nanostructuring can significantly reduce lattice thermal conductance by introducing interfaces, boundaries, and defects that scatter phonons more effectively than bulk materials. As a result, the mean free path of phonons decreases, leading to reduced thermal conductivity.[2,16]

Guo et al. conducted first-principles calculations to investigate the thermoelectric transport properties of SnSe.[17] Their results showed that for SnSe, the thermal conductivity is primarily influenced by phonons with short mean free paths (MFPs). At 300 K, phonons with MFPs ranging from 1 to 30 nm predominantly carry heat, with over 50% of them having wavelengths as short as 5 nm. At high temperatures, the phonon MFPs are further reduced to approx. 2nm. The phonon MFPs in SnSe are notably shorter than other well-known thermoelectric materials like PbSe and PbTe (up to~300 nm at 300 K),[18] accounting for the lower lattice thermal conductance in the former. In materials with dimensions larger than the phonon MFPs, heat transport follows a diffusive path instead of a ballistic one and adheres to Fourier's law of heat conduction.[19] For SnSe, the predominance of phonon modes with very short MFPs indicates that further reducing thermal conductivity through nanostructuring may not be feasible. To test this, we studied the thermal properties of SnSe nanoflakes at the nanoscale using atomic force microscopy (AFM)-based scanning thermal microscopy (SThM).

SThM is a technique that enables the measurement of local thermal conductance.[20] Operating as a variant of AFM, it employs a specialized probe with an integrated Wheatstone bridge circuit designed to map nanoscale thermal properties. Although experimental techniques like the opto-thermal Raman and micro-bridge methods reliably measure thermal conductivity, they face challenges when comparing the thermal conductivity of 2D materials with various layer numbers. Since the samples and the experimental conditions are different, directly comparing data from these techniques might be challenging.[21] SThM addresses this concern. While it cannot directly measure thermal conductivity, it can be used to study how thermal conductivity changes with the size of crystallites. Therefore, in this study, we employed AFM-based SThM measurements to examine the relationship between the thermal properties of SnSe nanoflakes and their thickness. [12]

**Methods:**

Thermolyne 46100 high-temperature furnace was used to grow 300 nm thick oxide layers on Silicon wafers using a dry oxygen method at 1200 °C. A Lambda Scientific LEOI-44 ellipsometer was employed to verify the thickness of the oxide layers. Before transferring SnSe nanoflakes, $SiO_2$ substrates were subjected to a sequential ultrasonication procedure, including acetone, isopropanol alcohol, and distilled water to remove any potential organic contaminates, followed by air drying. SnSe nanoflakes were mechanically exfoliated from bulk pieces (tin selenide, 99.999%, ThermoFisher Scientific) and transferred onto thermally grown $SiO_2$ substrates via scotch tape.

To investigate local thermal properties at the nanoscale, we utilized the Bruker NanoWizard-4XP series AFM, integrated with an Anasys Instruments VITA Module. All measurements were conducted under atmospheric conditions, acoustically and vibrationally isolated to minimize external interferences, and at room temperature to reflect practical applications. Additionally, we performed multiple scans to ensure consistency and reliability. SThM measurements were conducted in contact mode using the specialized VITA-HE-GLA-1 probe with a nominal resonance frequency of 50 kHz and a spring constant of 0.5 N/m. The topography and SThM maps were post-processed using Gwyddion, an open-source software.[22] Each image was processed using mean filtering to level the plane and align the rows, removing the background.

The probe acts as a resistive heater and is part of an integrated Wheatstone bridge circuit, as depicted in Figure 1b, allowing it to be heated to approximately 160 °C in active mode. Changes in the probe's electrical resistance can be detected as changes in voltage across the initially balanced bridge ($V_{out} = V_s - V_r$), generating thermal contrast maps alongside topographical data.

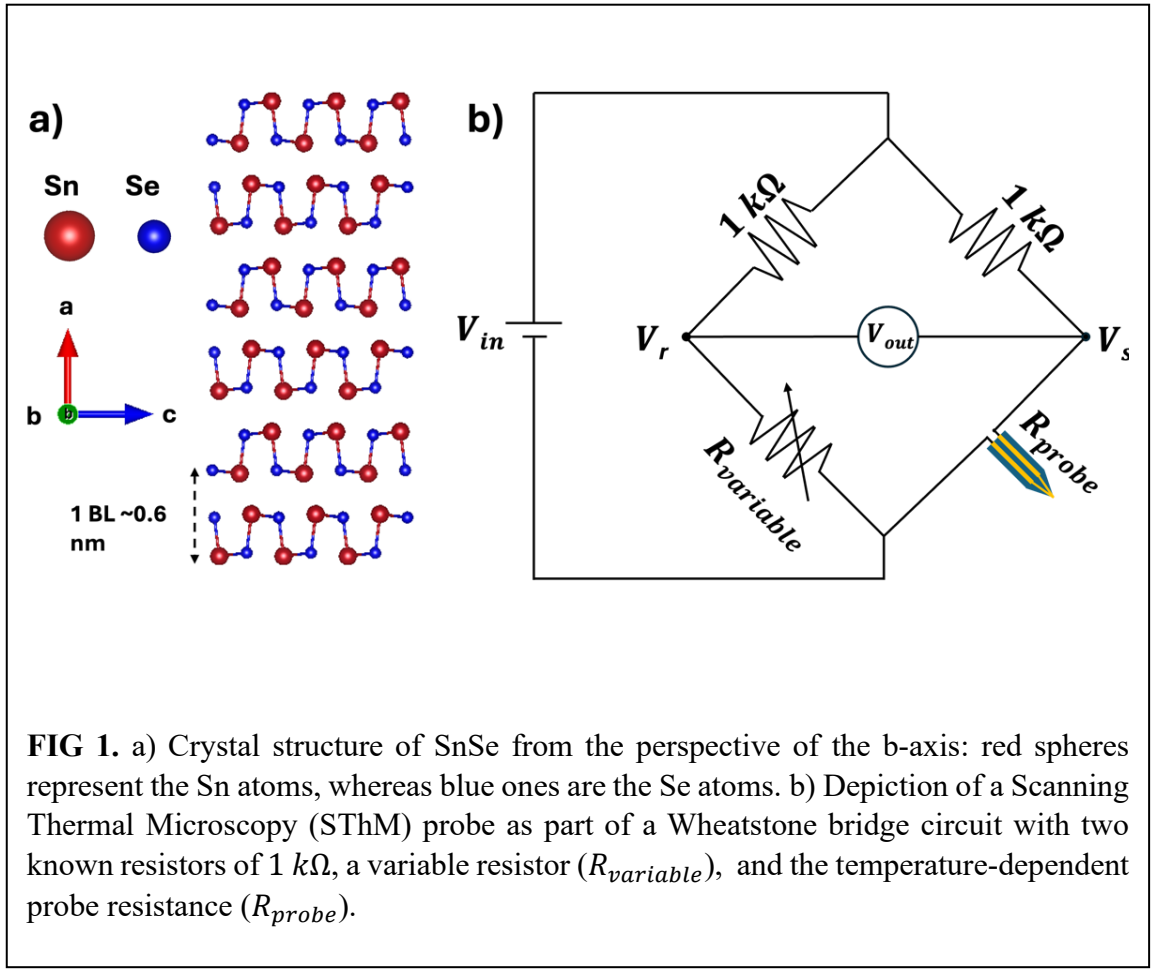

**FIG 1.** a) Crystal structure of SnSe from the perspective of the b-axis: red spheres represent the Sn atoms, whereas blue ones are the Se atoms. b) Depiction of a Scanning Thermal Microscopy (SThM) probe as part of a Wheatstone bridge circuit with two known resistors of 1 $k\Omega$, a variable resistor ($R_{variable}$), and the temperature-dependent probe resistance ($R_{probe}$).

### Results & Discussion

The SThM circuit records the voltage difference ($V_{out} = V_s - V_r$) across the initially balanced Wheatstone bridge ($V_s - V_r = 0$). As depicted in Figure 1b, the bridge configuration consists of two fixed resistors of 1 $k\Omega$, a variable resistor or potentiometer, and an SThM probe. Before the probe approached the surface, we applied a voltage of 1 V ($V_{in}$) to the Wheatstone bridge and then balanced the bridge in air, finding the probe's electrical resistance to be 415.4 $\Omega$. Once the probe contacted the SiO$_2$ surface, the probe's resistance decreased to 413.8 $\Omega$ due to the heat transfer from the probe to the surface. The bridge was rebalanced after contacting the sample, and the contact mode scanning was initiated.

Figure 2 shows topography and SThM maps taken on the SnSe nanoflakes. The topography map (Fig. 2a) and the vertical deflection map of the same area (Fig. 2b) show SnSe nanoflakes with varying thicknesses. In SThM measurements, regions with low thermal resistance allow efficient heat exchange between the probe and the sample, causing the probe to cool down. This results in a smaller voltage difference on the SThM map, represented by darker colors. Conversely, in areas with higher thermal resistance, heat transfer is not efficient, causing the SThM probe to retain more heat, which is then reflected as a higher voltage difference in the map and depicted with a brighter color.

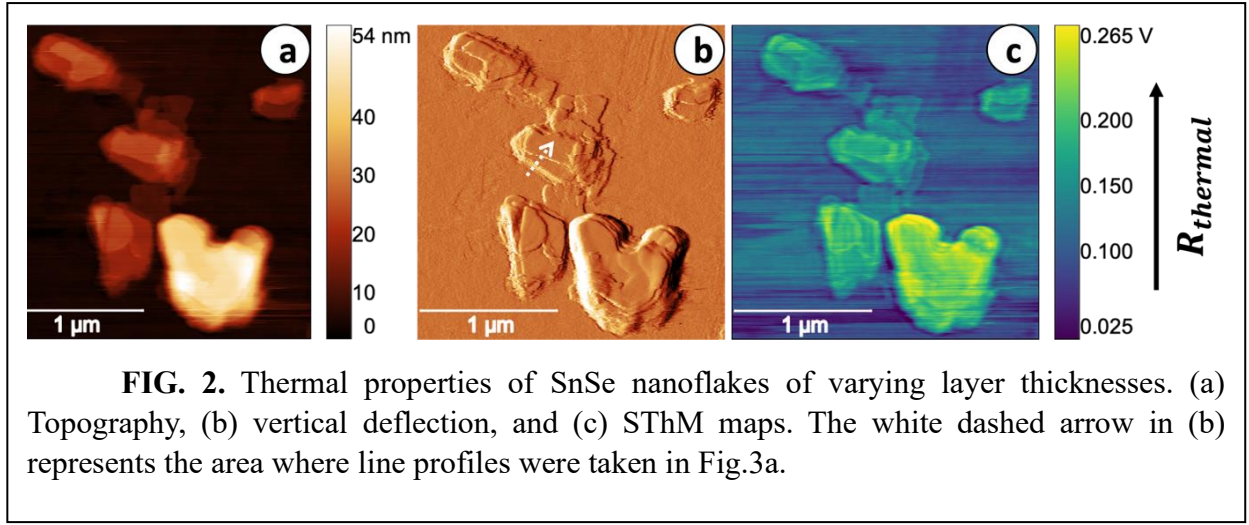

**FIG. 2.** Thermal properties of SnSe nanoflakes of varying layer thicknesses. (a) Topography, (b) vertical deflection, and (c) SThM maps. The white dashed arrow in (b) represents the area where line profiles were taken in Fig.3a.

**FIG. 3.** (a) A representative height and SThM voltage profile comparison. The line profiles were taken from the area (300 nm x 120 nm) shown in Fig.2b with a white-dashed arrow. (b) The relationship between the voltage drop measured in SThM and the varying thickness of SnSe nanoflakes. The orthogonal distance regression (ODR) fitting approach was performed to account for measurement errors in the voltage values.

The SThM map demonstrates a consistent behavior: as the layer thickness increases, so does the voltage drop across the Wheatstone bridge, which indicates increasing thermal resistance. This correlation is demonstrated in Figure 3a, where the height and SThM profiles were compared (from the area depicted by the white dashed arrow in Figure 2b). The topography line scan (blue curve) has two distinct steps corresponding to thicknesses of 11.5 nm (~ 19 layers) and 18.5 nm (~ 30 layers) of SnSe layers, respectively. The step edges of the flakes introduced some artifacts into SThM data, seen as sudden voltage jumps/drops on the SThM profile (red curve), but the overall trend of increasing thermal resistance with layer thickness is clear. To quantify the correlation between thickness and thermal resistance better, we analyzed SThM images of SnSe nanoflakes with various thicknesses and plotted the voltage drop against the layer thickness (Fig. 3b). Mean voltage drops and their corresponding standard deviations were extracted from the SThM map with the help of line profiles. The width of line was increased to ~120 nm to improve data accuracy while carefully excluding edge effects. The orthogonal distance regression (ODR) fitting approach was performed using MATLAB 2024a to minimize the weighted orthogonal distance between the measured SThM voltage and layer thickness, accounting for measurement errors in the voltage values. The result showed a linear trend, confirming that thermal resistance in SnSe nanoflakes increases linearly with thickness, as predicted by the Fourier law of heat conduction. It is anticipated that when the material dimensions exceed the phonon MFPs, frequent phonon scattering occurs. This results in heat transport following a diffusive path rather than a ballistic one, aligning with Fourier's law of heat conduction. In the case of SnSe, the prevalence of phonon modes with very short mean free paths suggests that further reducing thermal conductivity along the a-axis through nanostructuring is not feasible.

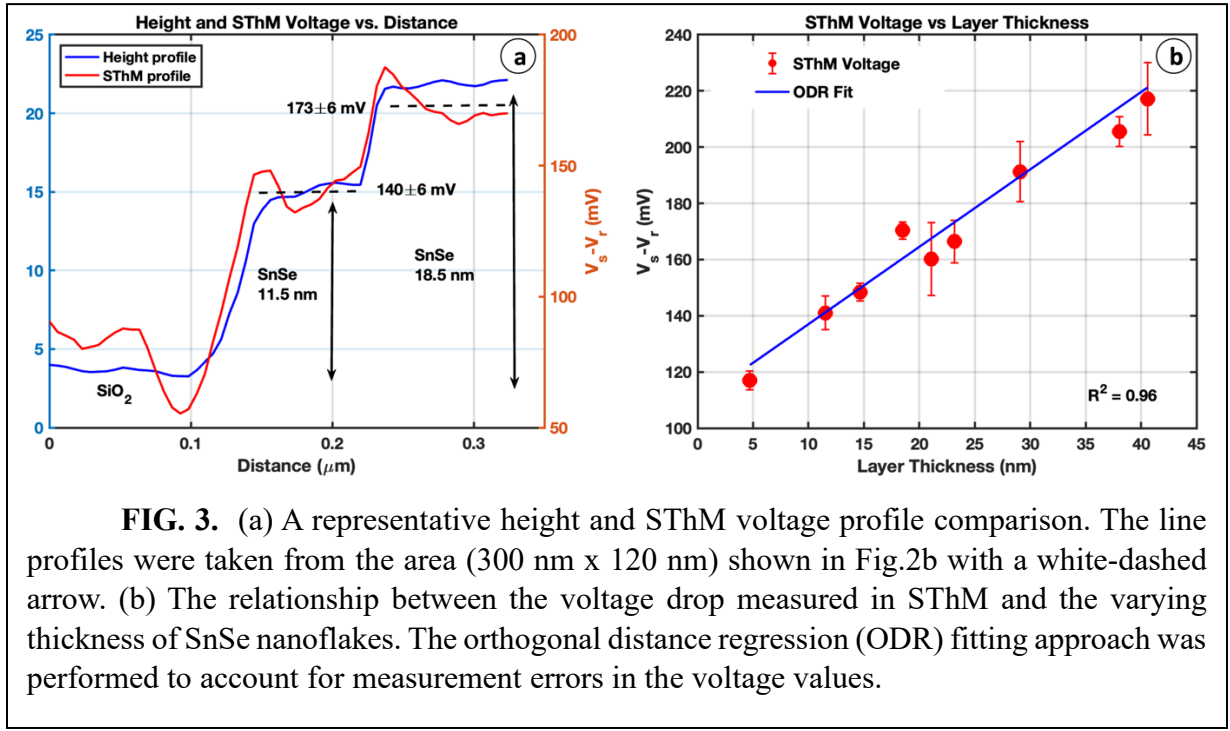

**FIG. 3.** (a) A representative height and SThM voltage profile comparison. The line profiles were taken from the area (300 nm x 120 nm) shown in Fig.2b with a white-dashed arrow. (b) The relationship between the voltage drop measured in SThM and the varying thickness of SnSe nanoflakes. The orthogonal distance regression (ODR) fitting approach was performed to account for measurement errors in the voltage values.

**Conclusion:**

We conducted AFM-based SThM measurements on mechanically exfoliated SnSe nanoflakes of varying thickness. Analysis of the SThM/AFM data revealed a linear relationship between the thickness of SnSe nanoflakes and the voltage drop, indicative of thermal resistance, across the Wheatstone bridge. In conclusion, our findings demonstrate that heat conduction in SnSe nanoflakes adheres to Fourier's law, due to the short phonon mean free path of approximately 5 nm at room temperature. This path is notably shorter than in other thermoelectric materials such as PbSe and PbTe, indicating that nanostructuring may not be a viable method for further reducing the thermal conductivity of SnSe along the out-of-plane (a-axis) direction.